\documentclass{article}

\usepackage[margin=1in]{geometry}
\usepackage{graphicx}
\usepackage{amsmath, amssymb}
\usepackage{booktabs}
\usepackage{multirow}
\usepackage{hyperref}
\usepackage{titlesec}
\usepackage{caption}
\usepackage{subcaption}
\usepackage{listings}
\usepackage{xcolor}
\usepackage{enumitem}
\usepackage{setspace}
\usepackage[switch]{lineno}
\usepackage{array}
\usepackage{algorithm}
\usepackage{algorithmic}  % or algorithmicx

\usepackage{tabularx}
\usepackage{array} % for >{\raggedright\arraybackslash}

\usepackage{tikz}
\usetikzlibrary{positioning,arrows.meta,fit,shapes,calc}

\usepackage[
  style=ieee,
  sorting=none
]{biblatex}
\title{ADAPT: AI-Driven Decentralized Adaptive Publishing Testbed}

\author{
  Md Motaleb Hossen Manik \\
  Department of Computer Science \\
  Rensselaer Polytechnic Institute \\
  Troy, New York 12180, USA
  \and
  Ge Wang \thanks{Corresponding author: Ge Wang, email: \texttt{wangg6@rpi.edu}} \\
  Department of Biomedical Engineering \\
  Rensselaer Polytechnic Institute \\
  Troy, New York 12180, USA
}

\date{}

\begin{document}
\maketitle

\begin{abstract}
Scholarly publishing faces increasingly strong stressors, including submission overload, reviewer fatigue, inconsistent evaluation, governance opacity, and vulnerability to manipulation in old and new forms. While recent studies applied artificial intelligence to improve specific steps (e.g., triage, reviewer recommendation, or automated critique), they typically work under centralized editorial control and offer limited mechanisms for system-level adaptivity and auditability.
Here we present \textbf{ADAPT} (AI-Driven Decentralized Adaptive Publishing Testbed), an agent-based environment that models journal management as a closed-loop control system rather than a fixed editorial workflow. ADAPT integrates interacting agents in various pools (authors, reviewers---human and AI---and rotating editors) coupled through policy-level control and diverse feedback signals. Governance adapts to backlog pressure, reviewer disagreement, paper quality drifting, and other relevant factors, while keeping human decision authority, role non-permanence, and data confidentiality.
We evaluate ADAPT in a discrete-time simulation setting across multiple operational regimes, including baseline operation, submission surges, quality drift, disagreement escalation, post-publication learning, and collusion suppression. Across regimes, we quantify backlog dynamics, reviewer load, coordination activity, and management performance. The results indicate that ADAPT works under nominal and perturbed conditions, exhibits bounded and interpretable responses under stress, and mitigates clusters with embedded interventions.
This feasibility demonstration suggests a promising direction of academic publishing practice, and can be extended to real-world implementations in suitable scenarios.
\end{abstract}


\section{Introduction}
\label{sec:intro}

Scholarly publishing is under major strains and positioned to meet new challenges. The growth in research data is exponential, and the volume of paper submissions has increased editorial workload and become a bottleneck, compromising decision latency and paper quality \cite{tennant2018state}. At the same time, many venues still rely on centralized editorial authority---often a small and relatively stable decision-making core---which creates scalability limits. This new trend interacts with long-standing issues such as inconsistent evaluation, bias, and lack of transparency in the editorial system \cite{lee2013bias}, demanding a novel solution to reduce operational cost, bridge the accountability gap, and upgrade the prevailing publishing models \cite{lee2013bias,tennant2018state}.

The above problems persist due to structural reasons: publishing behaves like a complex system. It comprises interacting agents (authors, reviewers, editors, institutions) whose behaviors co-evolve through delayed feedback, non-stationary input quality, and incentive-driven responses. Under such dynamics, local efficiency improvements do not necessarily yield global optimization. Review disagreement can rise near decision boundaries, overload can reduce review completeness, and incentives can produce emergent failure modes such as coordinated manipulations \cite{feliciani2019scoping,haug2015peer}. Yet governance in most venues remains comparatively static, adapting slowly and opaquely to changing situations. This mismatch motivates a different view of publishing: not as a centralized editorial workflow, but as a decentralized governance process that should remain interpretable, auditable, and robust under stress.

In this feasibility study, we introduce \textbf{ADAPT} (\textbf{A}I-\textbf{D}riven \textbf{D}ecentralized \textbf{A}daptive \textbf{P}ublishing \textbf{T}estbed), a framework that treats journal operation as a closed-loop governance system rather than a collection of fairly independent workflow elements. ADAPT monitors system-level signals---such as backlog pressure, reviewer disagreement, and review completeness---and updates a small set of bounded, interpretable policy variables (e.g., triage threshold, reviewer allocation, escalation sensitivity) while preserving protocol invariants including human authority, role non-permanence, and data confidentiality. ADAPT also incorporates post-publication outcomes as retrospective learning signals for credit updates across roles, enabling long-horizon incentive alignment without retroactively altering editorial decisions.

We evaluate ADAPT in a discrete-time, agent-based simulation setting across multiple operational regimes, including baseline operation, submission surges, quality drift, disagreement-driven uncertainty, post-publication learning, and collusion suppression. Across these regimes, the simulator enables controlled stress-testing of governance behavior: whether adaptation remains inactive under nominal conditions, responds coherently under sustained pressure, and recovers without oscillation when stress subsides.

Our main contributions are:
\begin{itemize}[leftmargin=1.2em]
    \item \textbf{A decentralized publishing testbed.} We propose ADAPT as a decentralized, feedback-driven testbed for studying policy-level AI-aided control in scholarly publishing, with rotating human roles and auditable governance actions.
    \item \textbf{Bounded, interpretable policy adaptation.} We formalize governance-level updates driven by aggregate system signals, avoiding manuscript-specific overrides to preserve procedural robustness, fairness and accountability \cite{rodden2019decentralized}.
    \item \textbf{Long-horizon learning via delayed outcomes.} We integrate post-publication impact as a retrospective signal for credit assignment across authors, reviewers, and governance roles, supporting incentive alignment over extended horizons \cite{liu2024enhancing}.
    \item \textbf{Simulation-based evaluation and ablations.} We evaluate stability and adaptability of our system under overload, drift, disagreement, and adversarial regimes, including mitigation ablations for collusion capture.
\end{itemize}

\paragraph{Paper organization.}
Section~\ref{sec:related_work} reviews related work.
Section~\ref{sec:methodology} presents the ADAPT framework, including design principles, stress models, system architecture, adaptive governance, incentive alignment, and auditability.
Section~\ref{sec:results} summarizes results across operational regimes.
Section~\ref{sec:discussion} discusses relevant issues.
Finally,
Section~\ref{sec:conclusion} concludes the paper.

\section{Related Work}
\label{sec:related_work}

This section reviews research threads that motivate ADAPT. We focus on (i) limitations of the conventional peer review workflow, (ii) feasibility of AI-assisted editorial pipelines, (iii) use and abuse of post-publication signals, and (iv) potential of decentralized governance that emphasizes auditability and incentive alignment. Together, these lines of work highlight why our ADAPT system is promising for future academic publishing practice.

\subsection{Traditional Peer Review}
Peer review has long been criticized for weak inter-reviewer agreement, vulnerability to bias,  and manipulations of various types \cite{smith2006peer,lee2013bias,haug2015peer}. Reforms such as open review and post-publication commentary can improve transparency, but they typically preserve centralized editorial structures and do not address workload burdens and long-horizon accountability \cite{tennant2017multi}.

\subsection{AI-Assisted Review and Editorial Support}
Recent studies explored AI for reviewer recommendation, triage, and decision support within existing editorial pipelines \cite{checco2021ai}. Other studies examined score calibration and related interventions that improve local consistency \cite{su2025icml2023rankingexperiment}. At the same time, concerns remain about encoding normative values and bias in automated systems \cite{birhane2022values}. Large language models have also been studied as simulated reviewers, but findings emphasize variability and prompt sensitivity, reinforcing that AI-aided review generation alone does not resolve system-level governance challenges \cite{jin2024agentreview,li2025llm}.

\subsection{Post-Publication Feedback}
Post-publication signals (e.g., bibliometric outcomes) provide feedback that can inform retrospective evaluation, but they are noisy and can be manipulated. Empirical work documented coercive citations and coordinated manipulations, motivating regulated use of bibliometric data as a constrained, auditable learning signal rather than a sole proxy for quality \cite{doi:10.1126/science.1212540,ibrahim2025citation}.

\subsection{Decentralized Science and Trust-Minimized Governance}
Decentralized science (DeSci) advocates protocol-based governance, transparency, and incentive alignment, but often leaves open how adaptive governance and human decision authority should be operationalized in day-to-day operations \cite{weidener2024decentralized}. Trust-minimized systems demonstrate that large-scale coordination can be achieved via auditable rules and incentives rather than reliance on centralized intermediaries \cite{nakamoto2008bitcoin}. This perspective motivates novel governance mechanisms that make policy evolution observable and resistant to capture.

% \paragraph{Connection to our work.}
Prior efforts either improve local workflow steps or propose decentralized coordination without a concrete, closed-loop control layer with human oversight. ADAPT bridges these directions by treating publishing as a policy-controlled governance system, where bounded interventions are triggered by explicit system signals and recorded as auditable events.

% ============================================================
\section{Methodology}
\label{sec:methodology}

ADAPT is a decentralized, adaptive, and auditable publishing framework, initially as a simulation testbed for proof of concept.
In ADAPT, “decentralized” refers to distributed governance constraints that reduce long-run concentration of authority: editor roles are modeled as non-permanent, interventions are triggered by aggregate signals, and governance actions are logged as auditable events.
The simulation does not aim to reproduce a specific journal’s staffing, but to test whether such protocol constraints can mitigate capture-like concentration patterns.
We define the system entities and observable signals, formalize policy-level governance control, specify the stochastic mechanics used to generate submissions and reviews, and describe logged outputs for reproducibility and auditability.
\textbf{All accept/reject outcomes in this paper are produced by a fixed rule; ADAPT’s contribution is a policy-level controller and an auditability layer.}

Across all regimes, adaptation operates at the \emph{policy level} (how the system allocates review effort and when it triggers structured interventions), rather than applying per-manuscript overrides. In any real deployment, AI will not make the final accept/reject decision, and ADAPT only recommends and logs policy-level actions, while the final outcome can be safely governed and ratified by human editors.

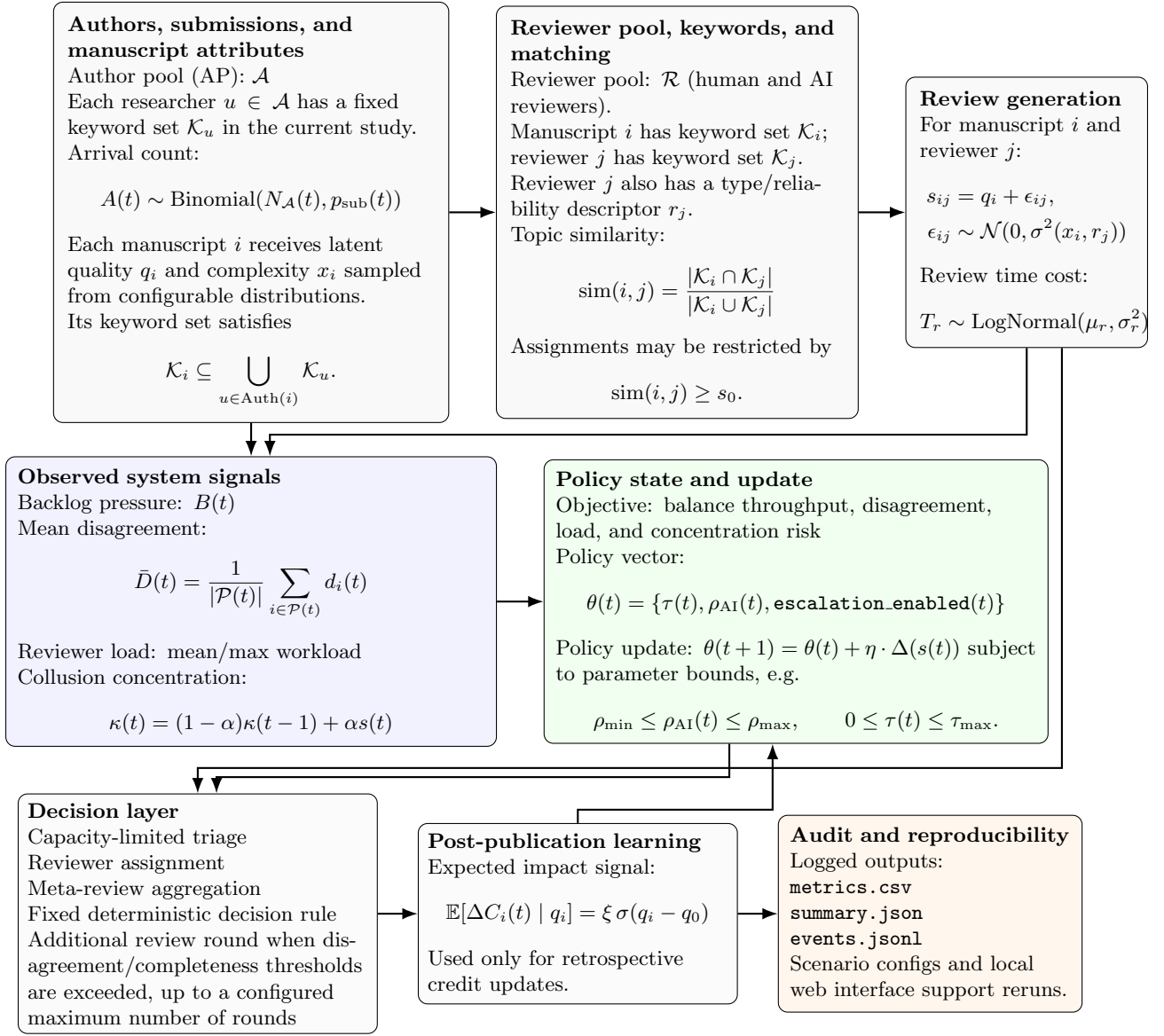
\begin{figure*}[!htb]
\centering
\begin{tikzpicture}[
    font=\small,
    >=Latex,
    box/.style={draw, rounded corners, align=left, inner sep=6pt, fill=gray!5},
    smallbox/.style={draw, rounded corners, align=left, inner sep=4pt, fill=gray!3},
    signal/.style={draw, rounded corners, align=left, inner sep=5pt, fill=blue!5},
    policy/.style={draw, rounded corners, align=left, inner sep=5pt, fill=green!6},
    audit/.style={draw, rounded corners, align=left, inner sep=5pt, fill=orange!7},
    arrow/.style={->, thick}
]
\footnotesize
% Row 1
\node[box, text width=55mm] (submissions) {
\textbf{Authors, submissions, and manuscript attributes}\\
Author pool (AP): $\mathcal A$\\
Each researcher $u\in\mathcal A$ has a fixed keyword set $\mathcal K_u$ in the current study.\\
Arrival count:
\[
A(t)\sim \mathrm{Binomial}(N_{\mathcal A}(t),p_{\text{sub}}(t))
\]
Each manuscript $i$ receives latent quality $q_i$ and complexity $x_i$ sampled from configurable distributions.\\
Its keyword set satisfies
\[
\mathcal K_i \subseteq \bigcup_{u\in \mathrm{Auth}(i)} \mathcal K_u .
\]
};

\node[box, text width=50mm, right=7mm of submissions] (matching) {
\textbf{Reviewer pool, keywords, and matching}\\
Reviewer pool: $\mathcal R$ (human and AI reviewers).\\
Manuscript $i$ has keyword set $\mathcal K_i$; reviewer $j$ has keyword set $\mathcal K_j$.\\
Reviewer $j$ also has a type/reliability descriptor $r_j$.\\
Topic similarity:
\[
\mathrm{sim}(i,j)=
\frac{|\mathcal K_i\cap \mathcal K_j|}
{|\mathcal K_i\cup \mathcal K_j|}
\]
Assignments may be restricted by
\[
\mathrm{sim}(i,j)\ge s_0 .
\]
};

\node[box, text width=32mm, right=7mm of matching] (review) {
\textbf{Review generation}\\
For manuscript $i$ and reviewer $j$:
\[
\begin{aligned}
s_{ij} &= q_i+\epsilon_{ij},\\
\epsilon_{ij} &\sim \mathcal N(0,\sigma^2(x_i,r_j))
\end{aligned}
\]
Review time cost:
\[
T_r\sim \mathrm{LogNormal}(\mu_r,\sigma_r^2)
\]
};

% Row 2
\node[signal, text width=70mm, below=5mm of submissions] (signals) {
\textbf{Observed system signals}\\
Backlog pressure: $B(t)$\\
Mean disagreement:
\[
\bar D(t)=\frac{1}{|\mathcal P(t)|}\sum_{i\in\mathcal P(t)} d_i(t)
\]
Reviewer load: mean/max workload\\
Collusion concentration:
\[
\kappa(t)=(1-\alpha)\kappa(t-1)+\alpha s(t)
\]
};

\node[policy, text width=72mm, right=7mm of signals] (policy) {
\textbf{Policy state and update}\\
Objective: balance throughput, disagreement, load, and concentration risk\\
Policy vector: \[
\theta(t)=\{\tau(t),\rho_{\text{AI}}(t),\texttt{escalation\_enabled}(t)\}
\]
Policy update: $\theta(t+1)=\theta(t)+\eta\cdot \Delta(s(t))$
subject to parameter bounds, e.g.
\[
\rho_{\min}\le \rho_{\text{AI}}(t)\le \rho_{\max},
\qquad
0\le \tau(t)\le \tau_{\max}.
\]
};

% Row 3
\node[smallbox, xshift = -8mm, text width=51mm, below=7mm of signals] (decision) {
\textbf{Decision layer}\\
Capacity-limited triage\\
Reviewer assignment\\
Meta-review aggregation\\
Fixed deterministic decision rule\\
Additional review round when disagreement/completeness thresholds are exceeded, up to a configured maximum number of rounds
};

\node[smallbox, text width=45mm, right=6mm of decision] (postpub) {
\textbf{Post-publication learning}\\
Expected impact signal:
\[
\mathbb E[\Delta C_i(t)\mid q_i]
=
\xi\,\sigma(q_i-q_0)
\]
Used only for retrospective credit updates.
};

\node[audit, text width=42mm, right=6mm of postpub] (audit) {
\textbf{Audit and reproducibility}\\
Logged outputs:\\
\texttt{metrics.csv}\\
\texttt{summary.json}\\
\texttt{events.jsonl}\\
Scenario configs and local web interface support reruns.
};

% Arrows row 1
\draw[arrow] (submissions.east) -- (matching.west);
\draw[arrow] (matching.east) -- (review.west);

% Row 1 to row 2
\draw[arrow] (submissions.south) -| (signals.north);
\draw[arrow] (review.south) -- ++(0,-1.3) -| ([xshift=8pt]signals.north);

% Row 2
\draw[arrow] (signals.east) -- (policy.west);

% Row 2 to row 3
\draw[arrow] ([xshift = -10mm]policy.south) -- ++(0,-0.5) -| ([xshift=8pt]decision.north);
\draw[arrow] ([xshift=15pt]review.south) -- ++(0,-6.3) -| (decision.north);

% Row 3
\draw[arrow] (decision.east) -- (postpub.west);
\draw[arrow] (postpub.east) -- (audit.west);

% Feedback
\draw[arrow] (postpub.north) -- ++(0,0.2) -| ([xshift=-10pt]policy.south);

\end{tikzpicture}
\caption{Unified graph of the default ADAPT simulator.}
\label{fig:unified_model_graph}
\end{figure*}

Figure~\ref{fig:unified_model_graph} summarizes the default ADAPT simulator instantiation used in this paper, linking the manuscript-level generative assumptions, observable governance signals, bounded policy updates, and delayed post-publication learning signals in one view. We next define the system entities and control variables used throughout the remainder of the methodology section.

% ============================================================
\subsection{Notations and System Entities}
\label{subsec:notation}

Time evolves in discrete time-steps $t = 0,1,\ldots,T-1$.
At each time-step, the system receives new manuscripts, selects a capacity-limited subset for review, produces review signals, and updates a small set of governance parameters.
Let $B(t)$ denote the backlog size (manuscripts awaiting processing) at time $t$.
Let $\mathcal{R}$ be the reviewer pool (human and AI agents), and let $\theta(t)$ denote the policy vector:
\begin{equation}
\theta(t) = \{\tau(t), \rho_{\text{AI}}(t), \texttt{escalation\_enabled}(t)\},
\label{eq:policy_vector}
\end{equation}
where $\tau(t)$ is the triage threshold, $\rho_{\text{AI}}(t)$ is the AI reviewer fraction used in reviewer sampling, and \\
\texttt{escalation\_enabled} is a Boolean flag controlling disagreement-driven escalation.

The testbed logs per-timestep metrics to \texttt{metrics.csv} (e.g., backlog, mean reviewer load, mean disagreement) and governance events to an append-only event log (Section~\ref{subsec:auditability}).

\paragraph{Modeling scope of the current testbed.}
The current ADAPT instantiation is manuscript-centered rather than participant-state-centered.
Specifically, the simulator assigns each manuscript a latent quality $q_i$ and complexity $x_i$, and models reviewer heterogeneity through reviewer type, workload, and noise characteristics.
We assume a pool of researchers that can supply authors and reviewers, and in the current study each researcher carries a fixed keyword set used for topical characterization.
For a manuscript with coauthor set $\mathrm{Auth}(i)$, the manuscript keyword set $\mathcal K_i$ is taken as a subset of the union of coauthor keyword sets, i.e.,
\[
\mathcal K_i \subseteq \bigcup_{u\in \mathrm{Auth}(i)} \mathcal K_u .
\]
Although real participants may act as authors, reviewers, and editors over longer horizons, the present paper does not introduce a single persistent participant-level quality variable shared across all such roles, and it does not model time-varying keyword profiles.
These simplifications are intentional: they keep the present testbed interpretable and reproducible while still allowing controlled studies of backlog pressure, disagreement, escalation, post-publication learning, and collusion mitigation.

% ============================================================
\subsection{Design Principles}
\label{subsec:vision_principles}

Unlike traditional editorial workflows that rely on centralized authority and static policies, ADAPT treats publishing operations as a closed-loop governance system.
In the proposed framework, AI assistance (e.g., triage support and signal aggregation) is embedded within policy updates that respond to system-level stress, while preserving protocol-level constraints.

ADAPT integrates three pillars:
\begin{enumerate}[leftmargin=1.2em]
    \item \textbf{Artificial intelligence} to scale triage assistance, reviewer allocation, and signal aggregation;
    \item \textbf{Decentralized governance} to reduce long-term concentration of authority through protocol constraints and transparent policies;
    \item \textbf{Credit dynamics} to align incentives with long-horizon outcomes rather than single decision outcomes.
\end{enumerate}

This system is implemented under the four design principles:

\paragraph{Decentralization and security.}
Authority is distributed across policies, participant pools, and protocol constraints, reducing single critical points of control.
Governance actions are externally auditable via an append-only record (Section~\ref{subsec:auditability}).

\paragraph{Adaptive governance.}
Governance parameters (e.g., $\tau(t)$ and $\rho_{\text{AI}}(t)$) evolve in response to measurable system signals.
Adaptation operates at the \emph{policy level}, supporting procedural consistency and avoiding manuscript-specific exceptions.

\paragraph{Incentive alignment.}
ADAPT rewards long-term impact and calibrated judgment.
Delayed post-publication feedback provides learning signals for credit evolution and slower-timescale governance adjustment (Section~\ref{subsec:credit_incentives}).

\paragraph{Human authority (deployment principle).}
In deployment, publication decisions and escalations are human-ratified, at least initially before super-human intelligence becomes reality in this domain.
In this paper, outcomes are generated by a deterministic rule to isolate governance effects from decision-model variation.

% ============================================================
\subsection{Stress Regimes}
\label{subsec:stress_models}

We evaluate ADAPT under stress regimes that reflect dominant journal-scale stressors and adversarial behaviors.
Each stressor is operationalized by an explicit, reproducible change in the editorial workflow:

\begin{itemize}[leftmargin=1.2em]
    \item \textbf{Submission overload:} changed arrival rate within a window, inducing backlog growth and reviewer fatigue \cite{SPITZER2026100276}.
    \item \textbf{Quality drift:} changed mean submission quality over time, challenging static triage and decision thresholds \cite{solomon2006peer}.
    \item \textbf{Reviewer disagreement (epistemic uncertainty):} increased review noise that raises disagreement signals and triggers structured escalation when enabled.
    \item \textbf{Collusion / cluster capture:} a coordinated subset increases within-cluster co-review concentration; a detection rule triggers decentralization/rotation mitigation unless disabled (ablation).
\end{itemize}

These stressors motivate adaptive, system-level governance rather than static editorial management, and will be examined empirically in Section~\ref{sec:results}.

% ============================================================
\subsection{System Architecture}
\label{subsec:architecture}
\begin{figure*}[t]
    \centering
    \includegraphics[width=0.65\linewidth]{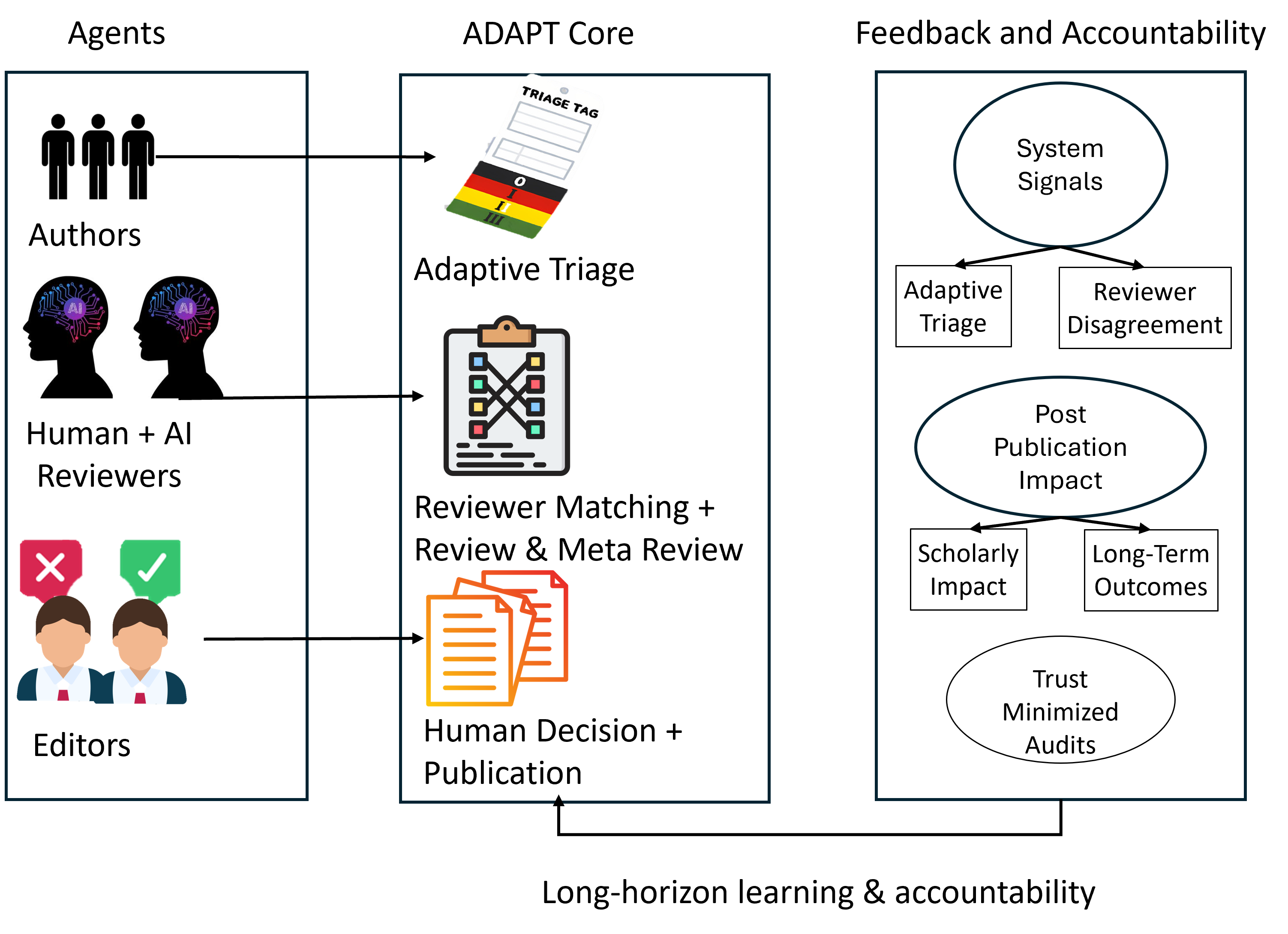}
\caption{ADAPT overview with agents, governance, feedback, and auditability.}    
\label{fig:architecture}
\end{figure*}

The unified graph in Fig.~\ref{fig:unified_model_graph} summarizes the simulator’s default variables and control relations, while Fig.~\ref{fig:architecture} presents the corresponding end-to-end workflow architecture.
The framework ADAPT is an end-to-end publishing system composed of (i) submissions and triage, (ii) reviewer assignment and review generation, (iii) meta-review signal aggregation, (iv) deterministic decision, (v) policy update, and (vi) auditable logging (Fig.~\ref{fig:architecture}).

\subsubsection{Per-timestep Workflow}
\label{subsec:state_machine}

Algorithm~\ref{alg:adapt_loop} summarizes the ADAPT workflow.

\begin{algorithm}[t]
\caption{ADAPT per-timestep loop (testbed)}
\label{alg:adapt_loop}
\begin{algorithmic}[1]
\STATE Receive new submissions; append to backlog.
\STATE Select a capacity-limited set via triage using threshold $\tau(t)$.
\STATE Assign $k$ reviewers using an AI fraction $\rho_{\text{AI}}(t)$ and workload constraints.
\STATE Generate reviews; aggregate into meta-review signals (disagreement, completeness).
\STATE If escalation is enabled and signals exceed thresholds, add a bounded additional round (max rounds).
\STATE Decide outcome via a fixed deterministic rule (accept/reject/revise) based on the aggregated reviews.
\STATE Update policy $\theta(t)\rightarrow\theta(t{+}1)$ using backlog and disagreement signals.
\STATE Log governance events to an append-only audit trail.
\end{algorithmic}
\end{algorithm}

\subsubsection{Agent Pools and Latent Variables}
\label{subsec:agent_pools_arch}

In the testbed, each manuscript $i$ has latent \emph{quality} $q_i$ and \emph{complexity} $x_i$ sampled from configurable distributions.
Reviewers are partitioned into human and AI pools with workload counters and heterogeneous noise characteristics.

A simple instantiation consistent with our simulator is:
\begin{equation}
s_{ij} = q_i + \epsilon_{ij},
\qquad
\epsilon_{ij} \sim \mathcal{N}\!\big(0,\ \sigma^2(x_i,r_j)\big),
\label{eq:review_score_model}
\end{equation}
where the noise variance increases with manuscript complexity and depends on the reviewer descriptor $r_j$, which in the current testbed captures reviewer type/reliability.
The disagreement spike is generated via an explicit noise multiplier.

\subsubsection{Meta-review Signals}
\label{subsec:meta_signals}

For each processed manuscript $i$, meta-review computes a disagreement signal $d_i(t)$ and a completeness signal $c_i(t)$ from the set of collected reviews.
We report the system-level mean disagreement as:
\begin{equation}
\bar{D}(t) = \frac{1}{|\mathcal{P}(t)|}\sum_{i\in\mathcal{P}(t)} d_i(t),
\label{eq:mean_disagreement}
\end{equation}
where $\mathcal{P}(t)$ denotes manuscripts processed at timestep $t$.
Analogously, completeness may be summarized as $\bar{Q}(t)$ from $\{c_i(t)\}$ when enabled.

% ============================================================
\subsection{Adaptive Governance}
\label{subsec:governance_control}

ADAPT updates governance through bounded, interpretable policy changes driven by aggregate signals.
The objective is to manage throughput and stabilize quality under stress.

\subsubsection{Governance objective and constraints}
\label{subsec:governance_objective}

ADAPT can be viewed as a constrained governance-control problem.
At each timestep, the controller seeks to improve editorial throughput and review quality proxies while limiting overload, disagreement, and concentration risk.
A compact stylized objective is:
\begin{equation}
U(t)=
w_B\,(-B(t))
+
w_D\,(-\bar D(t))
+
w_L\,(-L(t))
+
w_C\,(-\kappa(t))
+
w_P\,P(t),
\label{eq:governance_objective}
\end{equation}
where $B(t)$ is backlog pressure, $\bar D(t)$ is mean disagreement, $L(t)$ is reviewer-load pressure, $\kappa(t)$ is the concentration metric for capture-like clustering, and $P(t)$ denotes a positive performance term such as timely processing or publication-quality proxy.
The precise weights are journal-dependent and need not be fixed universally.

The controller operates subject to explicit operational constraints.
In the present testbed, these include bounded policy variables (e.g., $\rho_{\min}\le\rho_{\text{AI}}(t)\le\rho_{\max}$ and $0\le\tau(t)\le\tau_{\max}$), capacity limits on review effort per timestep, bounded escalation rounds, and the deployment principle that final publication decisions remain human-ratified.
More generally, a journal may impose service-level constraints such as an upper bound on mean review-cycle duration or a limit on acceptable reviewer workload.
Within ADAPT, such requirements can be encoded either through the objective weights or through explicit hard constraints in the policy update rule.

\subsubsection{Observable System Signals}
\label{subsec:signals}

The ADAPT controller monitors:
\paragraph{Backlog pressure.} $B(t)$, the number of manuscripts awaiting processing.

\paragraph{Disagreement.} $\bar{D}(t)$, the mean disagreement computed from per-manuscript meta-review signals (Eq.~\eqref{eq:mean_disagreement}).

\paragraph{Reviewer load.} Mean and max reviewer workloads summarize utilization and help diagnose capacity stress.

\paragraph{Concentration (collusion).} A concentration metric $\kappa(t)$ derived from within-cluster co-review share (Section~\ref{subsec:collusion_metric}) detects cluster capture, which may suggest potential collusion.

\subsubsection{Control Variables}
\label{subsec:control_vars}

ADAPT expresses governance through a small number of interpretable variables (Eq.~\eqref{eq:policy_vector}):

\paragraph{Triage threshold $\tau(t)$.}
Controls selectivity under capacity constraints; higher $\tau$ prioritizes decision reliability at the expense of throughput.

\paragraph{AI reviewer fraction $\rho_{\text{AI}}(t)$.}
Controls the mixture of AI vs.\ human reviewers in assignment; increasing $\rho_{\text{AI}}$ expands review capacity with a policy-level transparency.

\paragraph{Escalation enablement.}
When enabled, disagreement/completeness thresholds trigger a bounded additional review round (max rounds), converting uncertainty into targeted adjudication.

\subsubsection{Policy Update}
\label{subsec:objective_update}

Policies update via bounded steps:
\begin{equation}
\theta(t{+}1) = \theta(t) + \eta \cdot \Delta\!\big(s(t)\big),
\label{eq:policy_update}
\end{equation}
where $s(t)$ denotes observed signals and $\Delta(\cdot)$ is an interpretable rule.
The update rule $\Delta(\cdot)$ should be interpreted as a bounded controller that heuristically improves the governance objective in Eq.~\eqref{eq:governance_objective} while respecting operational constraints.
In the testbed, updates are constrained by caps (e.g., $\rho_{\text{AI}}\in[\rho_{\min},\rho_{\max}]$, $0\le\\tau\le\tau_{\max}$) and may include hysteresis to reduce oscillations.
Importantly, adaptation changes \emph{how} reviews are allocated and escalated, not \emph{which} manuscripts are accepted through manuscript-specific exceptions.
The parameter \texttt{triage\_step} sets the controller’s step size for $\tau(t)$: smaller values adjust triage more gradually (reducing saturation and oscillation risk), while larger values react more aggressively and can drive $\tau(t)$ to the cap $\tau_{\max}$ under sustained stress.
We report a sensitivity sweep over \texttt{triage\_step} in the disagreement-spike regime in Subsection~\ref{subsec:results_epistemic} (Table~\ref{tab:triage_step_sweep}).
%============================================================
\subsection{Generative Models and Core Mechanics}
\label{subsec:stochastic_models_params}

This subsection specifies the stochastic mechanics used to generate submissions and review signals in the simulation testbed.

\subsubsection{Manuscript arrivals and latent attributes}
Let $A(t)$ be the number of new submissions at timestep $t$.
We model arrivals with a simple stochastic process (e.g., Binomial or Poisson approximation for a sufficiently large author pool):
\begin{equation}
A(t) \sim \mathrm{Binomial}\!\big(N_{\mathcal{A}}(t),\, p_{\text{sub}}(t)\big),
\label{eq:arrivals_binomial}
\end{equation}
where $N_{\mathcal{A}}(t)$ is the active author pool and $p_{\text{sub}}(t)$ is the submission probability (optionally perturbed by scenario overrides).
Each manuscript receives latent quality $q_i$ and complexity $x_i$ sampled from configurable distributions, and complexity increases review noise.

\subsubsection{Reviewer workload and completion cost}
At each timestep, reviewers are sampled subject to workload limits.
To capture skewed time cost, we model review time cost using a log-normal form:
\begin{equation}
T_r \sim \mathrm{LogNormal}(\mu_r,\sigma_r^2),
\label{eq:review_time_lognormal}
\end{equation}
which induces variable per-timestep processing cost and contributes to backlog growth when demand exceeds capacity.

\subsubsection{Review signals, disagreement, and escalation}
For each processed manuscript, the simulator samples $k$ reviewers and generates noisy review signals.
Meta-review aggregates these signals into a disagreement statistic and other summary quantities used by policy control.
A disagreement-spike is implemented by increasing the review noise via an explicit multiplier during a defined window.
Escalation, when enabled, adds a bounded additional round of reviews when disagreement exceeds a threshold (up to a configured max rounds).

\subsubsection{Post-publication signal}
For long-horizon learning experiments, published manuscripts produce a synthetic impact signal that is positively related to latent quality.
A minimal instantiation is:
\begin{equation}
\mathbb{E}\!\left[\Delta C_i(t)\mid q_i\right] = \xi \cdot \sigma\!\big(q_i - q_0\big),
\label{eq:impact_growth}
\end{equation}
with stochastic realization (e.g., Poisson sampling).
This signal is used \emph{only} for retrospective credit updates (not retroactive decision changes).

For clarity, Table~\ref{tab:variables_distributions} summarizes the default variables, formulas, distributional choices, and rationale used in the current ADAPT simulator instantiation.
\begin{table*}[t]
\centering
\footnotesize
\caption{Default ADAPT simulator variables, distributions, and rationale.}
\label{tab:variables_distributions}
\begin{tabularx}{\textwidth}{p{0.85cm} p{2cm} p{2.75cm} p{2.7cm} p{6.3cm}}
\toprule
\textbf{Object} & \textbf{Meaning} & \textbf{Default formulation in this paper} & \textbf{Distribution / rule} & \textbf{Why this default is used here} \\
\midrule
$A(t)$ & New submissions at timestep $t$ & Eq.~\eqref{eq:arrivals_binomial} & Binomial count process & A simple arrival model for timestep-wise submissions; easy to perturb under surges and replace in later journal-specific versions. \\

$q_i$ & Manuscript latent quality & Sampled per manuscript & Configurable latent draw & Provides a latent quality variable for simulation; not assumed directly observable. \\

$x_i$ & Manuscript complexity & Sampled per manuscript & Configurable latent draw & Lets review noise and effort depend on manuscript difficulty in a simple, interpretable way. \\

$s_{ij}$ & Review score for manuscript $i$ by reviewer $j$ & Eq.~\eqref{eq:review_score_model} & Additive noisy score model & Transparently links latent quality to reviewer observations while allowing disagreement through reviewer-dependent noise. \\

$\epsilon_{ij}$ & Review noise & Part of Eq.~\eqref{eq:review_score_model} & Gaussian noise scaled by $x_i$ and reviewer reliability & A simple uncertainty model for reviewer variation; disagreement spikes are modeled by increasing this noise. \\

$T_r$ & Review time cost & Eq.~\eqref{eq:review_time_lognormal} & Log-normal & Captures positive, skewed processing cost without a more complex task-time model. \\

$\bar D(t)$ & Mean disagreement signal & Eq.~\eqref{eq:mean_disagreement} & Aggregated from processed manuscripts & Provides a compact observable signal for epistemic stress and escalation control. \\

$\theta(t)$ & Governance state & Eq.~\eqref{eq:policy_vector} & Policy vector & Limits adaptation to a small set of interpretable control variables. \\

$\theta(t{+}1)$ & Policy update & Eq.~\eqref{eq:policy_update} & Bounded rule-based update & Supports adaptive yet auditable policy changes while avoiding manuscript-specific exceptions. \\

$\Delta C_i(t)$ & Post-publication impact increment & Eq.~\eqref{eq:impact_growth} & Mean linked to latent quality, with stochastic realization & Serves as a delayed, noisy proxy for retrospective learning only, not retroactive decision revision. \\

$\kappa(t)$ & Concentration metric for collusion / capture & Eq.~\eqref{eq:concentration_ema} & Exponentially smoothed within-cluster share & Gives a simple, auditable signal for concentration growth and mitigation triggers. \\

\begin{tabular}[c]{@{}c@{}}$S_a(t)$\\$S_r(t)$\end{tabular} &
\begin{tabular}[c]{@{}l@{}}Author and\\reviewer credit\end{tabular} &
\begin{tabular}[c]{@{}l@{}}Eqs.~\eqref{eq:author_credit_update}\\and \eqref{eq:reviewer_credit_update}\end{tabular} &
Incremental update rules &
Lets delayed outcomes shape incentives over time while keeping decision-time governance separate from ex post credit assignment. \\
\bottomrule
\end{tabularx}
\end{table*}

% ============================================================
\subsection{Keyword-Based Manuscript--Reviewer Matching}
\label{subsec:keyword_matching}

To instantiate topical alignment, we use a stylized keyword model.
Let $\mathcal{K}$ be a keyword universe with $|\mathcal{K}| = K$.
Each researcher $u$ is associated with a keyword set $\mathcal{K}_u \subseteq \mathcal{K}$, which is fixed in the current study.
For manuscript $i$ with coauthor set $\mathrm{Auth}(i)$, the manuscript keyword set satisfies
\[
\mathcal{K}_i \subseteq \bigcup_{u\in \mathrm{Auth}(i)} \mathcal{K}_u .
\]
Each reviewer $j$ has a keyword set $\mathcal{K}_j \subseteq \mathcal{K}$.
Similarity is computed via Jaccard overlap:
\[
\mathrm{sim}(i,j) = \frac{|\mathcal{K}_i \cap \mathcal{K}_j|}{|\mathcal{K}_i \cup \mathcal{K}_j|}.
\]
Assignments may be restricted by a threshold $\mathrm{sim}(i,j)\ge s_0$, which reduces review noise by improving topical alignment.
Future extensions may allow researcher keyword profiles to evolve over time, but the present study keeps them fixed for clarity and reproducibility.
% ============================================================
\subsection{Collusion / Cluster Detection and Mitigation}
\label{subsec:collusion_metric}

To stress-test adversarial attacks, we simulate a coordinated reviewer subset that increases within-cluster co-review share $s(t)$.
We define an exponentially smoothed concentration metric:
\begin{equation}
\kappa(t) = (1-\alpha)\,\kappa(t-1) + \alpha \cdot s(t),
\label{eq:concentration_ema}
\end{equation}
with smoothing parameter $\alpha$.
Detection triggers an intervention if $\kappa(t)$ exceeds a threshold for a specified patience window.

\paragraph{Mitigation action.}
When intervention is active, mitigation reduces within-cluster share through a decay term (controlled by a mitigation strength parameter), representing diversification/rotation actions at the policy level.
%\paragraph{Mitigation ablation.}
We implement a configuration switch (\texttt{disable\_capture\_mitigation}) that prevents intervention, enabling a comparison where concentration remains elevated when governance response is disabled (Section~\ref{subsec:results_collusion}).

% ============================================================
\subsection{Credit Dynamics and Incentive Alignment}
\label{subsec:credit_incentives}

ADAPT maintains dynamic credit for authors and reviewers to align incentives with long-horizon outcomes.
In simulation, post-publication impact is modeled as a synthetic citation-like signal derived from latent quality, and is used \emph{only} for retrospective credit updates (no retroactive decision change).

\subsubsection{Post-publication Impact Signal}
\label{subsec:postpub_signal}

Let $C_i$ denote the realized impact of manuscript $i$ after publication (simulation-defined). Since such an impact is noisy and potentially gameable in real systems, ADAPT treats it as a constrained feedback channel (e.g., via smoothing and consistency checks) rather than a direct proxy for quality.

\subsubsection{Author Credit Updates}
\label{subsec:author_credit}

Each author maintains credit $S_a(t)$ updated by deviation from a nominal expected impact $\bar{C}$:
\begin{equation}
S_a(t{+}1) = S_a(t) + \alpha_a \cdot \big(C_i - \bar{C}\big).
\label{eq:author_credit_update}
\end{equation}

\subsubsection{Reviewer Credit Updates}
\label{subsec:reviewer_credit}

Reviewer credit reflects calibration between ex ante assessments and ex post outcomes:
\begin{equation}
S_r(t{+}1) = S_r(t) + \alpha_r \cdot \phi_r\!\big(R_{r,i},\, C_i\big),
\label{eq:reviewer_credit_update}
\end{equation}
where $R_{r,i}$ is the reviewer assessment and $\phi_r(\cdot)$ is an alignment function (e.g., sign agreement or a bounded loss on prediction error).

% ============================================================
\subsection{Auditability and Trust-Minimization}
\label{subsec:auditability}

ADAPT decouples accountability from content exposure by logging governance actions (not manuscript content) as auditable events.
In the testbed, an append-only event stream records triage statistics, policy updates, and regime-specific signals (e.g., collusion state), while excluding manuscript text and identities.
Logged events include: triage summaries (backlog before/after, processed count), policy updates ($\rho_{\text{AI}}(t)$, $\tau(t)$, \texttt{escalation\_enabled}), and in the collusion regime, concentration and intervention flags.
The audit layer is implementation-agnostic: deployments may use write-once logs, institutional ledgers, or distributed commitment registries, provided append-only and tamper-evident properties hold.

% ============================================================
\subsection{Observed Metrics, Baselines, and Reproducibility}
\label{subsec:governance_metrics}

The simulator writes a per-timestep time series (\texttt{metrics.csv}) and an append-only event log (\texttt{events.jsonl}).
The primary observed metrics include the backlog $B(t)$, mean disagreement $\bar{D}(t)$, reviewer load, and policy state $\{\rho_{\text{AI}}(t), \tau(t), \texttt{escalation\_enabled}(t)\}$.
For the collusion regime, we log the within-cluster share $s(t)$, concentration $\kappa(t)$, and intervention flag.
Unless otherwise specified, all experiments share identical baseline parameters summarized in Table~\ref{tab:baseline_params}.
Each regime differs only by a scenario configuration, enabling fair comparison.

\begin{table*}[t]
\centering
\caption{Baseline simulation parameters and default configuration.}
\label{tab:baseline_params}
\begin{tabular}{p{2.8cm} p{12.5cm}}
\hline
Category & Values \\
\hline
Submissions & mean arrivals $=30$/timestep; quality $\mu=0.6$, $\sigma=0.15$; complexity $\mu=0.5$, $\sigma=0.2$ \\
Reviewers & humans=200; AI pool=30; max\_load=6; ai\_enabled=True \\
Review process & $k=3$; \textbf{ai\_fraction\_target (initial)}: baseline $=0.2$; stress scenarios set 0.1 unless noted; escalation=True; disc\_th=1.4; comp\_th=0.55 \\
Capacity/Triage & max\_reviews/timestep=180; triage\_th0=0.45 \\
Decision & accept\_th=0.7; reject\_th=0.4 \\
Governance & backlog\_high=40; backlog\_low=10; ai\_min=0.1; ai\_max=0.6; ai\_step=0.05; triage\_step=0.03; triage\_max=0.7 \\
\hline
\end{tabular}
\end{table*}

\noindent\textbf{Note on reported policy values.}
Table~\ref{tab:baseline_params} reports baseline configuration settings, while all policy values discussed in Section~\ref{sec:results} (e.g., $\rho_{\text{AI}}(t)$ and $\tau(t)$) are taken from the realized time series in \texttt{metrics.csv}.
Scenario configurations may override initial values and policy caps. Hence, we report figure- and table-level policy numbers from the corresponding run directories.

\paragraph{Reproducibility.}
Each run produces \texttt{metrics.csv}, \texttt{summary.json}, and an append-only event log under a run directory.
Scenario configurations are specified in \texttt{configs/scenarios/*.yaml}, enabling reproducible regeneration of the figures and tables reported in Section~\ref{sec:results}.
In addition to command-line reruns, we provide a lightweight local web interface that allows a reader to select a figure panel, modify exposed protocol parameters, and regenerate the corresponding output from the underlying scripts.
The current code, configurations, and local web interface are available at \url{https://github.com/manikm-114/ADAPT_2}.
We additionally performed parameter-sensitivity sweeps by rerunning scenarios with single-parameter overrides (e.g., governance step sizes), and report only those sweep results whose override values are recorded in the run provenance.

\paragraph{Proxy indicators (validity).}
The quantities reported in \texttt{metrics.csv} are used as \emph{operational proxy indicators} inside a stylized simulator, not as universally validated domain metrics.
Each metric maps to an intuitive construct needed for governance control: backlog $B(t)$ measures throughput pressure, mean disagreement $\bar{D}(t)$ captures epistemic uncertainty in aggregated reviews, reviewer load reflects capacity stress, escalation counts measure the frequency of targeted adjudication under uncertainty, and the concentration metric $\kappa(t)$ approximates capture-like clustering in reviewer assignments.
We assess \emph{face validity} through expected-direction behavior under controlled regimes (e.g., surges increase backlog; disagreement spikes increase escalations; capture regimes increase $\kappa(t)$), and we report multi-seed robustness and parameter-sensitivity sweeps where appropriate.
Accordingly, we interpret these metrics as internally consistent signals for comparing governance responses across regimes, and we avoid claiming external construct validity beyond the simulation setting.

% ============================================================
\section{Numerical Results}
\label{sec:results}

We report our initial feasibility results across operational regimes focusing on (i) stability and backlog control, (ii) governance activation patterns (e.g., $\rho_{\text{AI}}$ and $\tau$),
(iii) uncertainty handling via disagreement and escalation, (iv) long-horizon learning driven by post-publication signals, and (v) robustness to collusion/cluster capture.
Figures~\ref{fig:baseline_surge}--\ref{fig:postpub_governance} summarize time-series behavior, and Fig.~\ref{fig:collusion_capture} summarizes capture mitigation.

% ------------------------------------------------------------
\subsection{Baseline Stability and Surge Recovery}
\label{subsec:results_baseline_surge}

\begin{figure*}[h]
  \centering
  \begin{subfigure}[t]{0.48\linewidth}
    \centering
    \includegraphics[width=\linewidth]{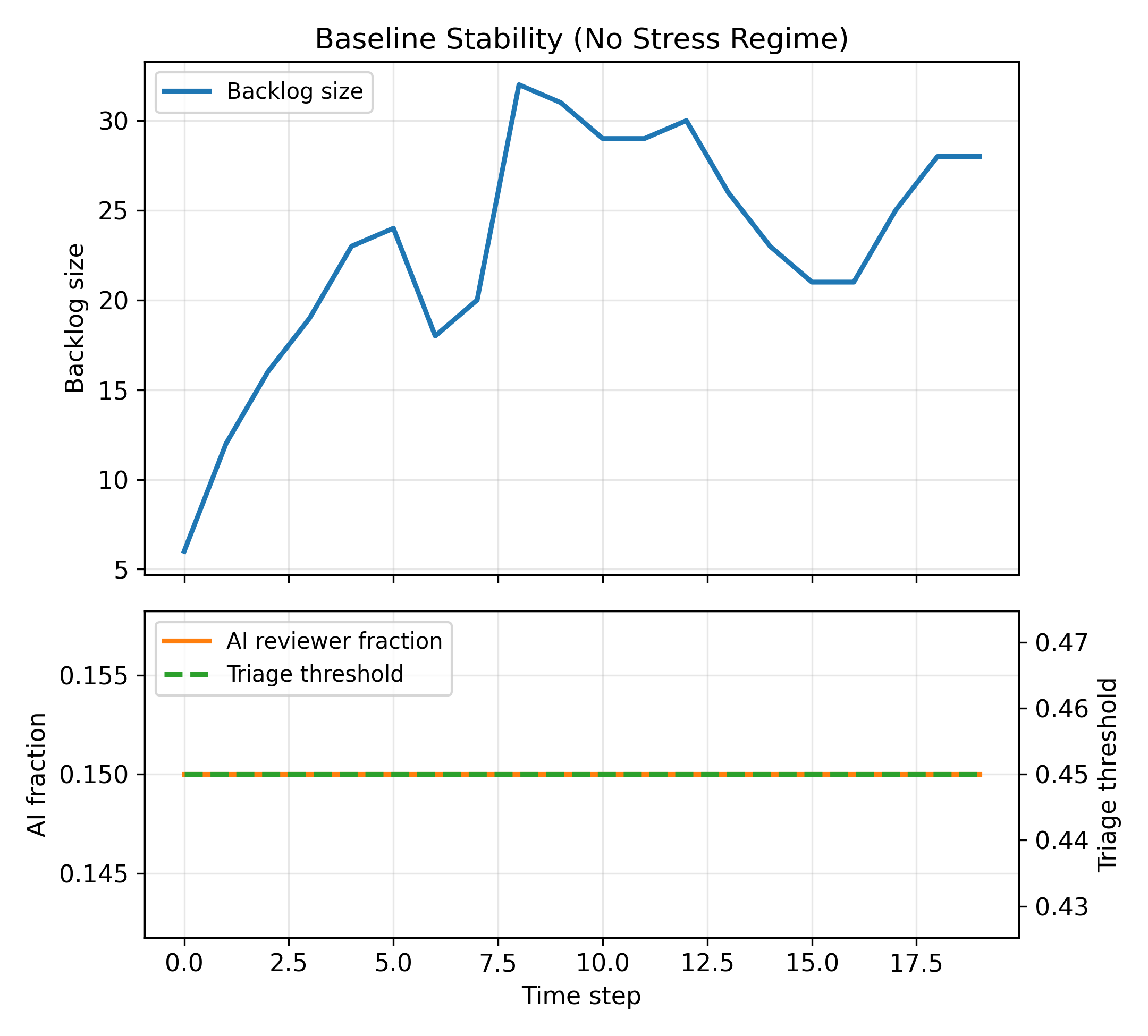}
    \caption{Baseline stability}
    \label{fig:baseline_surge_a}
  \end{subfigure}
  \hfill
  \begin{subfigure}[t]{0.48\linewidth}
    \centering
    \includegraphics[width=\linewidth]{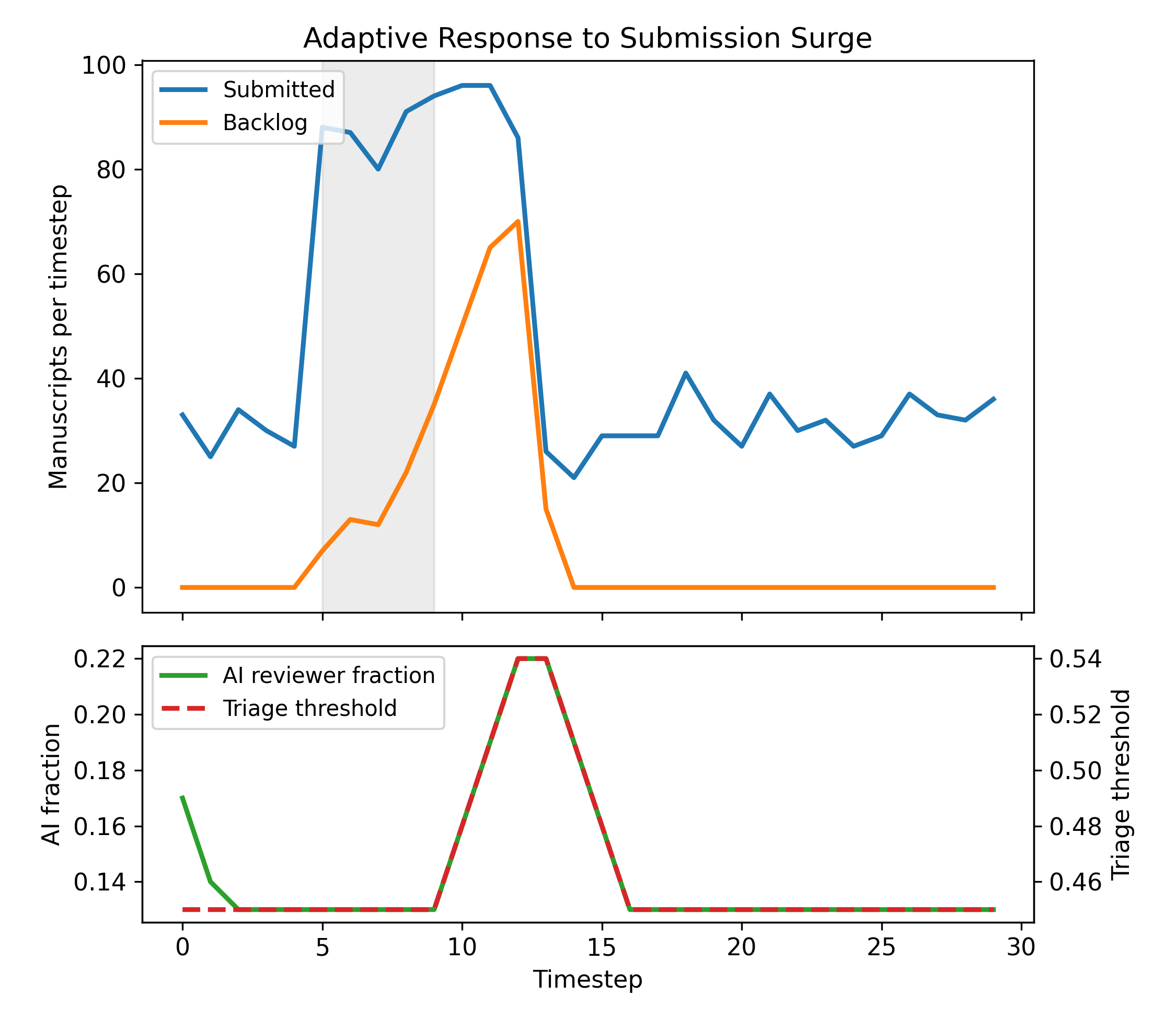}
    \caption{Submission surge recovery}
    \label{fig:baseline_surge_b}
  \end{subfigure}
  \caption{ADAPT system behavior: (a) Baseline operation; (b) Submission surge recovery.}
  \label{fig:baseline_surge}
\end{figure*}

Under nominal operation (Fig.~\ref{fig:baseline_surge}a), backlog remains bounded and governance remains effectively inactive:
policy parameters do not exhibit sustained drift in the absence of persistent stress signals.
In the submission surge regime (Fig.~\ref{fig:baseline_surge}b), increased arrivals push demand above capacity, causing backlog growth.
ADAPT responds with bounded policy adjustments and backlog recovery, after which policies relax rather than remaining saturated.

% ------------------------------------------------------------
\subsection{Epistemic Stress: Quality Drift and Disagreement}
\label{subsec:results_epistemic}

\begin{figure*}[!htb]
  \centering
  \begin{subfigure}[t]{0.48\linewidth}
    \centering
    \includegraphics[width=\linewidth]{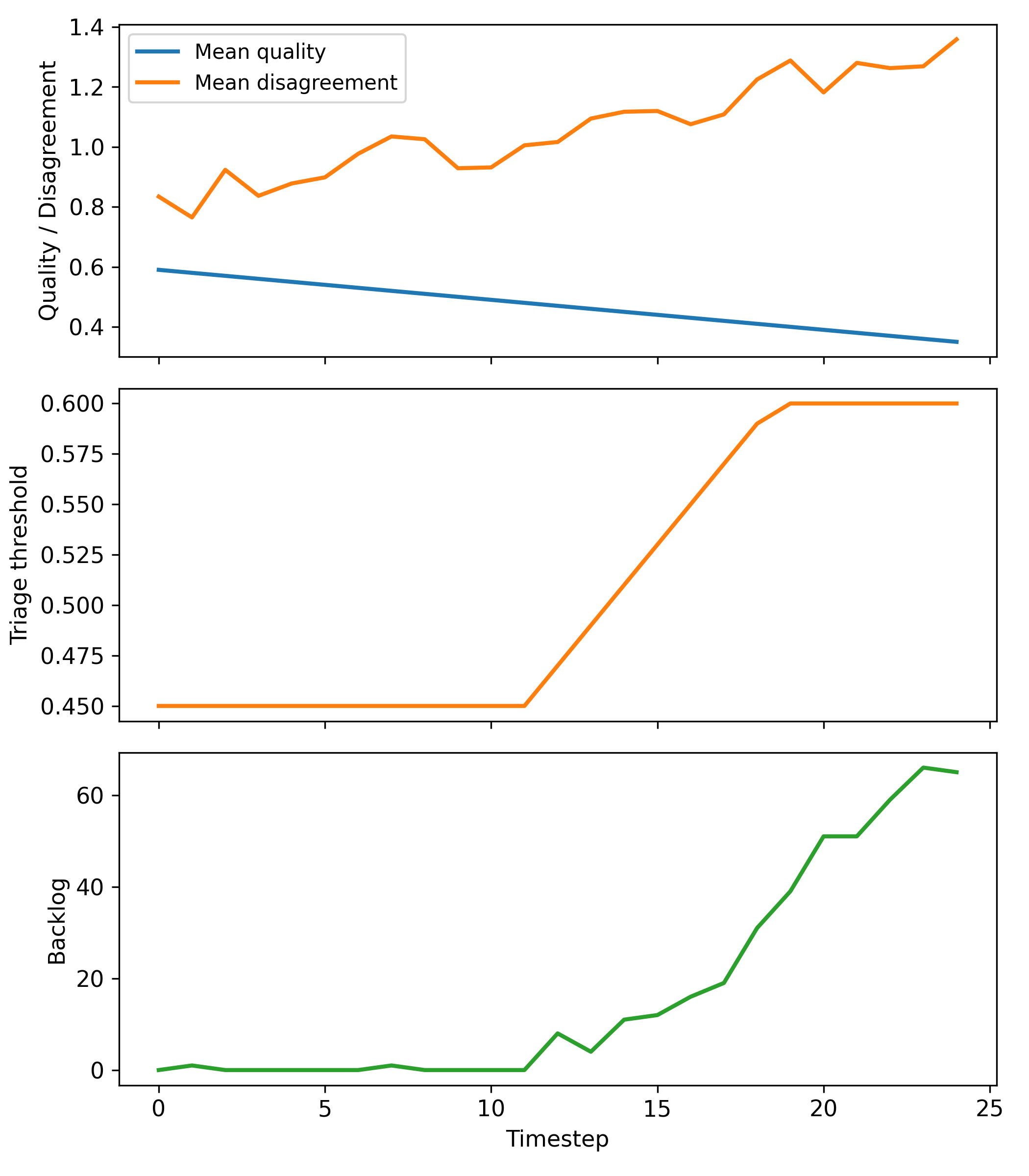}
    \caption{Quality drift}
    \label{fig:epistemic_stress_a}
  \end{subfigure}
  \hfill
  \begin{subfigure}[t]{0.48\linewidth}
    \centering
    \includegraphics[width=\linewidth]{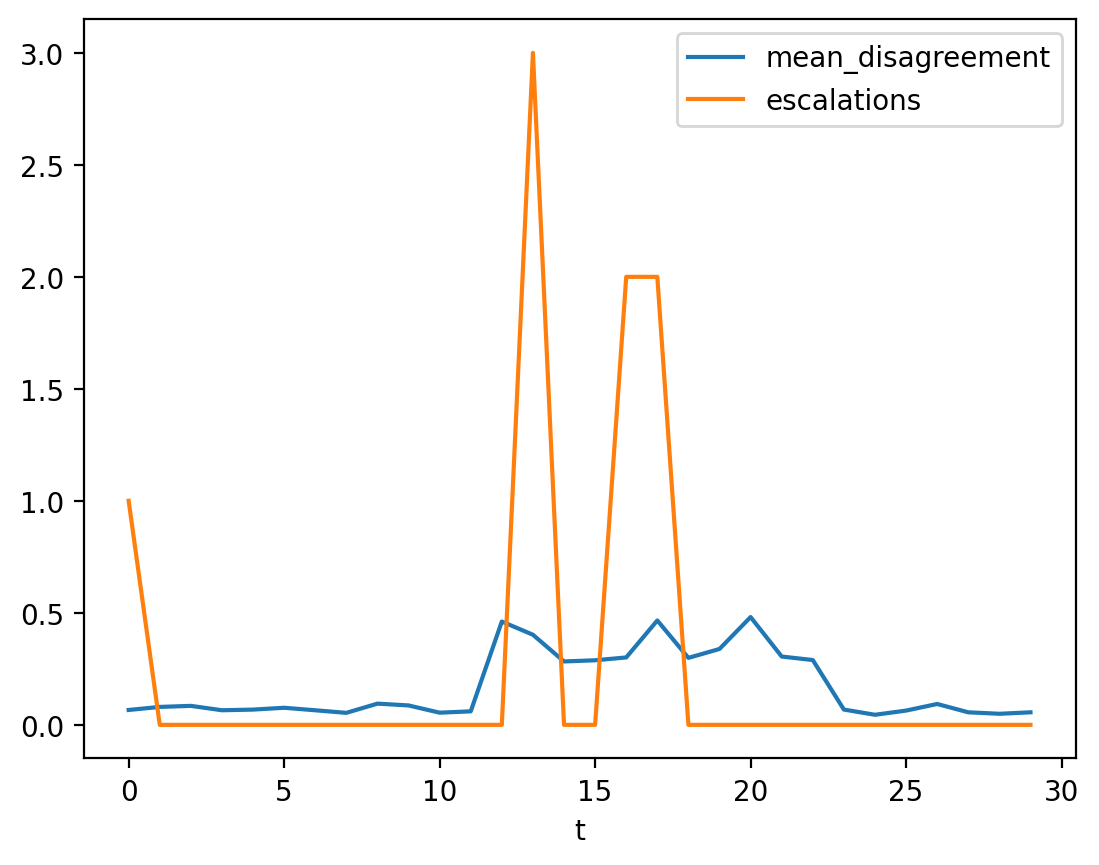}
    \caption{Disagreement spike}
    \label{fig:epistemic_stress_b}
  \end{subfigure}
  \caption{Epistemic stress regimes: (a) Quality drift; (b) Disagreement spike.}
  \label{fig:epistemic_stress}
\end{figure*}

\begin{figure}[t]
  \centering
  \includegraphics[width=\linewidth]{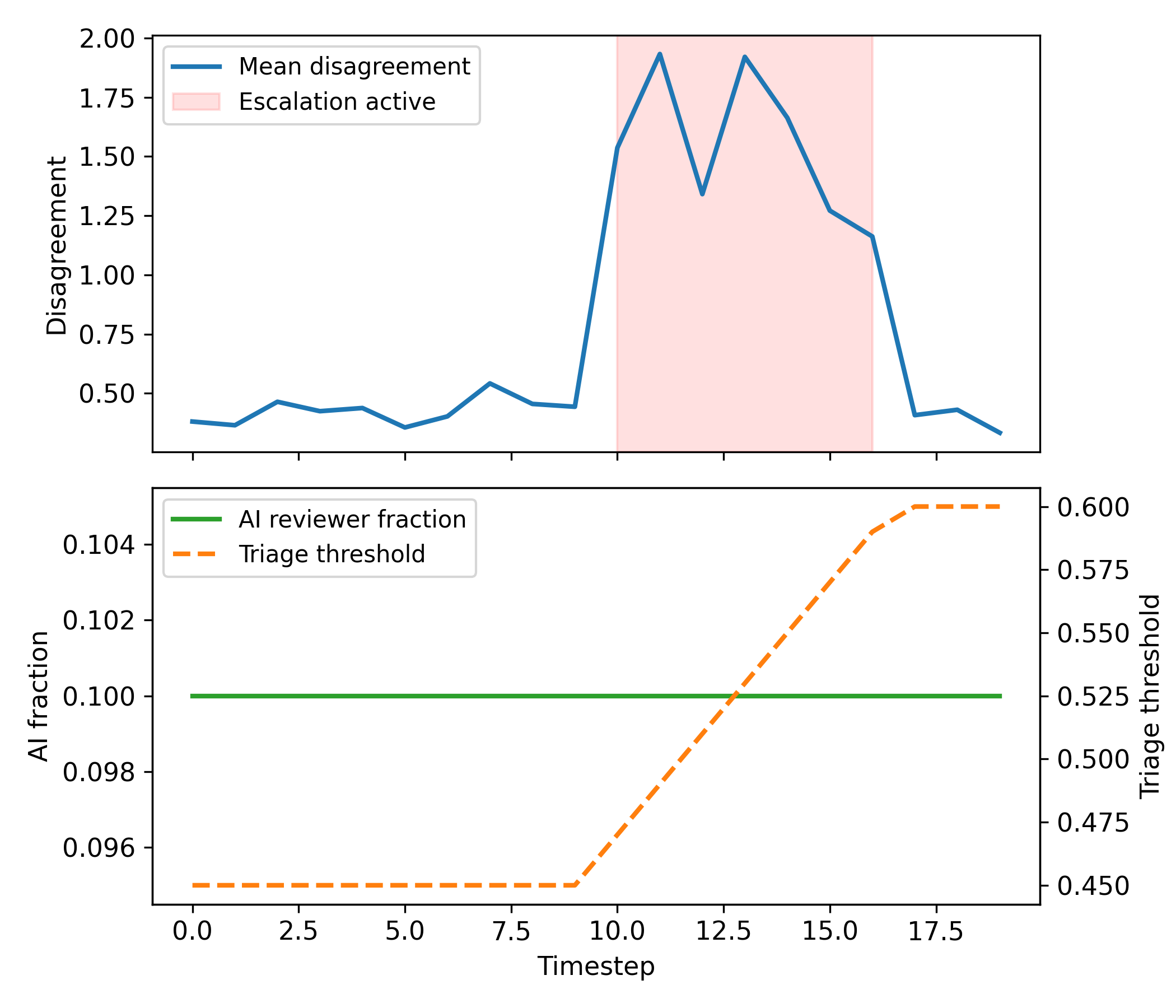}
  \caption{Disagreement-driven response in the spike regime.}
  \label{fig:disagreement_governance}
\end{figure}

Figure~\ref{fig:epistemic_stress} summarizes two epistemic stress regimes.
Under quality drift (Fig.~\ref{fig:epistemic_stress}a), declining input quality increases uncertainty near the decision boundary and induces a bounded increase in triage selectivity.
Under a disagreement spike (Fig.~\ref{fig:epistemic_stress}b), uncertainty is driven primarily by increased reviewer variance; the testbed triggers escalation events during the spike window and responds with bounded policy adjustments (Fig.~\ref{fig:disagreement_governance}).
Although escalation is disabled at the final timestep in this representative run, escalation events occur earlier during the spike window, consistent with a time-varying policy state.
In this run, the testbed logs $12$ escalation events in total, with a maximum of $4$ escalations in a single timestep.

\paragraph{Multi-seed robustness (disagreement spike).}
To assess robustness, we reran the disagreement-spike regime over $10$ random seeds.
Across runs, the final backlog is \textbf{median $182$} (min $160$, max $197$).
The final triage threshold saturates at its cap $\tau(T){=}0.70$ in all seeds, while the final AI reviewer fraction varies with realized trajectories: \textbf{median $0.475$} (min $0.450$, max $0.600$).
Escalation activity is consistently triggered during the spike window, with \textbf{median total escalations $11$} (min $6$, max $14$) and \textbf{median max escalations per timestep $4$} (min $2$, max $5$).

(For reference, the representative disagreement-spike run used in Table~\ref{tab:final_policy_states} ends with backlog $160$ and $\rho_{\mathrm{AI}}(T)=0.45$.)

\paragraph{Sensitivity to controller step size (\texttt{triage\_step}).}
To probe how controller aggressiveness affects recovery under epistemic stress, we swept the triage update step size (\texttt{governance.triage\_step}) in the disagreement-spike regime (fixed seed $123$), summarized in Table~\ref{tab:triage_step_sweep}.
With a smaller step of $0.01$, the system leaves the final triage threshold below its cap ($\tau(T){=}0.67$) and achieves the lowest residual backlog (final backlog $=96$), but with higher escalation activity (total escalations $=14$).
For larger steps ($0.02$--$0.10$), the policy saturates at $\tau(T){=}0.70$ and residual backlog is higher (final backlog $185$--$229$) with fewer total escalations ($5$--$12$).
This highlights a tradeoff between more gradual triage adaptation (lower backlog) and escalation burden under sustained uncertainty.

\begin{table}[t]
\centering
\small
\caption{Disagreement spike sensitivity to \texttt{triage\_step} (seed $123$).}
\label{tab:triage_step_sweep}
\begin{tabular}{c c c c}
\toprule
triage\_step & Final backlog & $\tau(T)$ & Total escal. \\
\midrule
0.01 & 96  & 0.67 & 14 \\
0.02 & 185 & 0.70 & 5  \\
0.03 & 210 & 0.70 & 12 \\
0.05 & 229 & 0.70 & 6  \\
0.07 & 215 & 0.70 & 7  \\
0.10 & 214 & 0.70 & 7  \\
\bottomrule
\end{tabular}
\end{table}

\subsection{Final-step Policy State Across Regimes}
\label{subsec:results_policy_table}

Table~\ref{tab:final_policy_states} reports the final-step governance state (last timestep) for the representative runs used to generate the main figures, including whether escalation was enabled and whether the collusion intervention activated.
Multi-seed robustness results (including disagreement-spike medians and ranges) are reported separately in Section~\ref{subsec:results_epistemic}.

\begin{table*}[h]
\centering
\small
\caption{\textbf{Final-step governance state by regime.}}

\label{tab:final_policy_states}
\begin{tabular}{p{4.5cm}p{2.25cm}p{2.3cm}p{1.2cm}p{2.1cm}p{1.5cm}}
\toprule
Scenario & AI fraction & Triage threshold & Escalation enabled & Final backlog & Intervention first $t$ \\
\midrule
Baseline & 0.15 & 0.45 & No & 26 & -- \\
Submission surge & 0.13 & 0.45 & No & 0 & -- \\
Quality drift & 0.10 & 0.60 & No & 59 & -- \\
Disagreement spike & 0.45 & 0.70 & No & 160 & -- \\
Collusion (mitigation enabled) & 0.60 (0.10--0.60) & 0.70 (0.45--0.70) & No & 133.5 (18--166) & 11 (10--13) \\
Collusion (mitigation disabled) & 0.60 (0.15--0.60) & 0.70 (0.48--0.70) & No & 140 (27--195) & -- \\
\bottomrule
\end{tabular}
\end{table*}

% ------------------------------------------------------------
\subsection{Post-Publication Learning}
\label{subsec:results_postpub}

\begin{figure*}[!htb]
  \centering
  \begin{subfigure}[t]{0.48\linewidth}
    \centering
    \includegraphics[width=\linewidth]{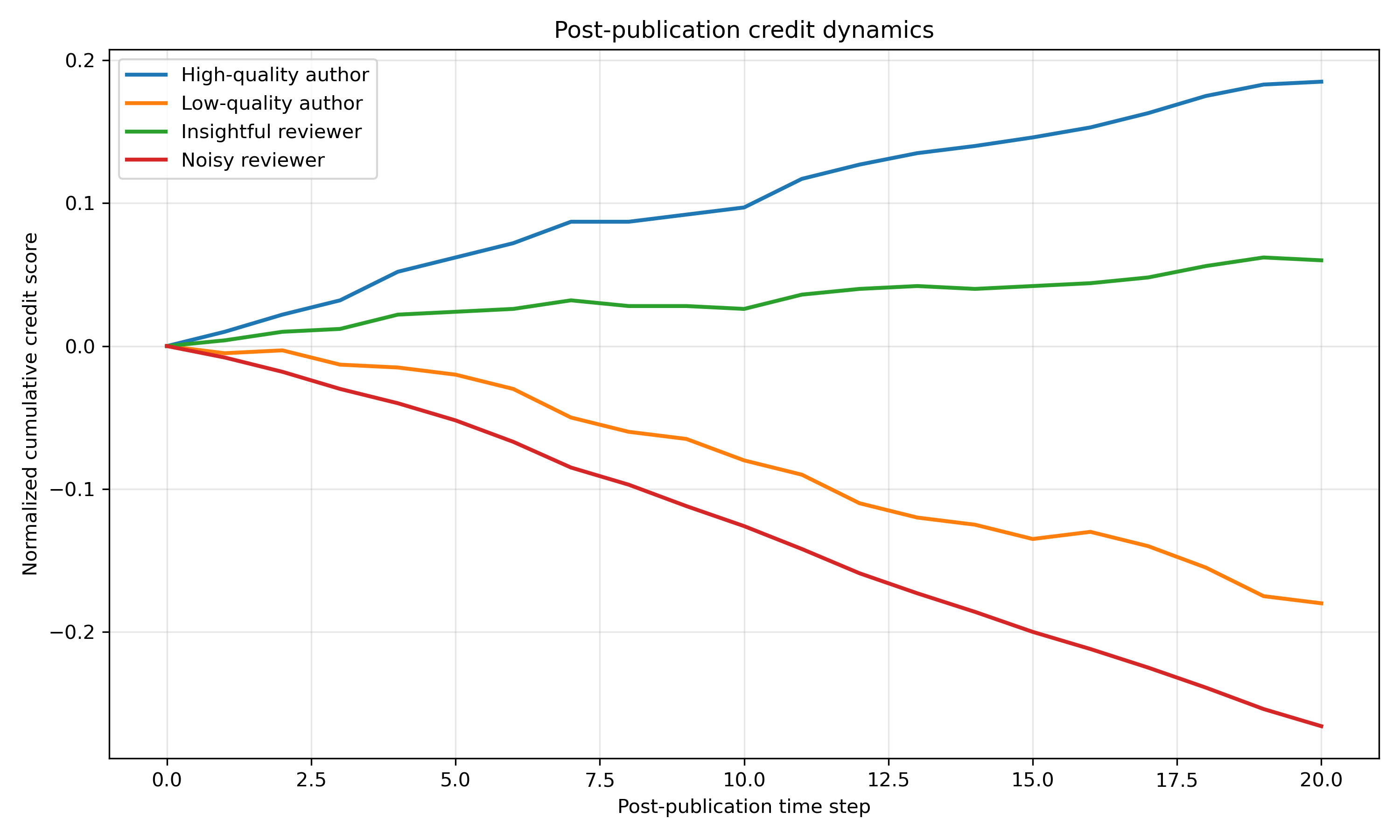}
    \caption{Post-publication credit dynamics}
    \label{fig:postpub_governance_a}
  \end{subfigure}
  \hfill
  \begin{subfigure}[t]{0.48\linewidth}
    \centering
    \includegraphics[width=\linewidth]{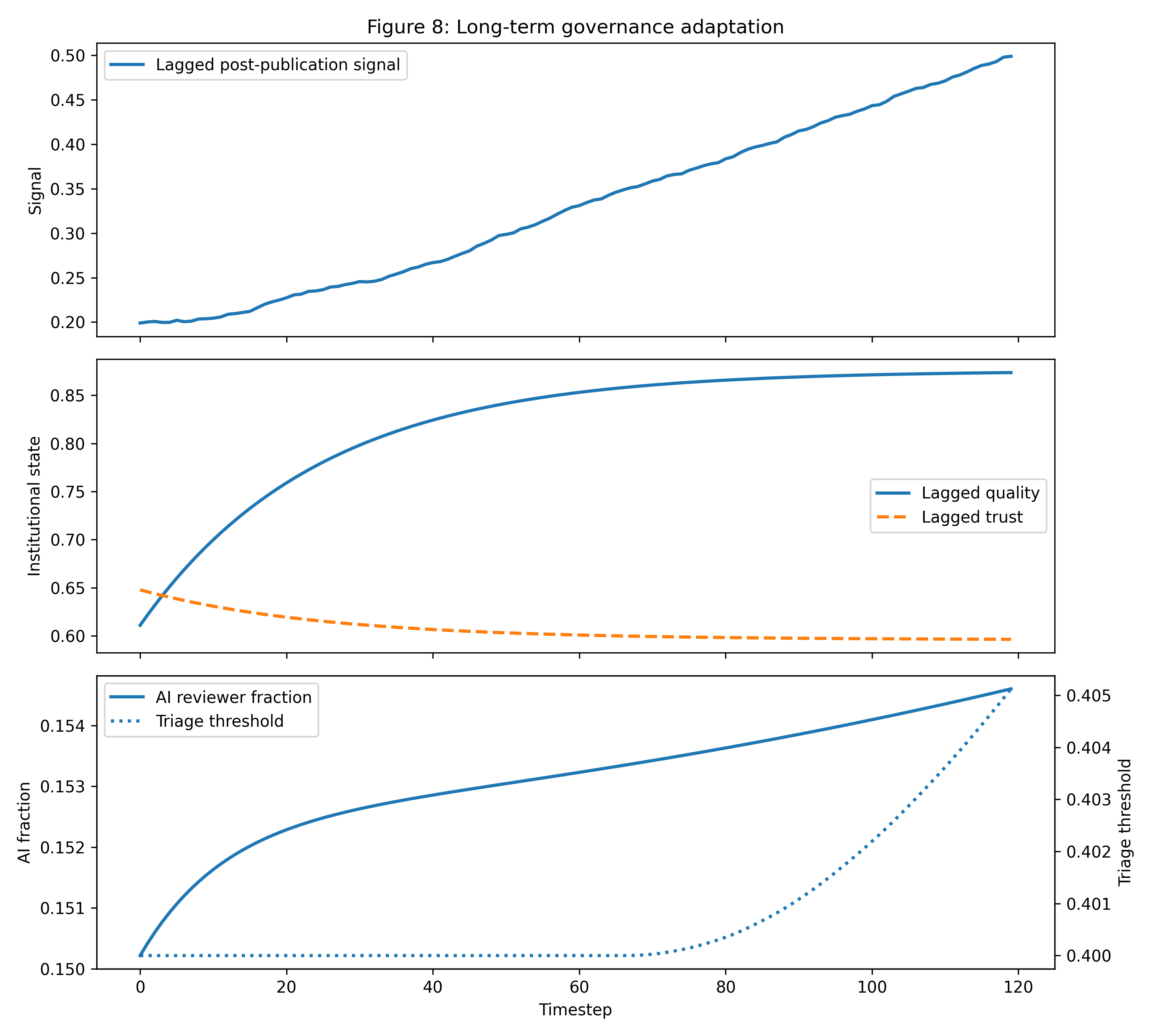}
    \caption{Long-term governance adaptation}
    \label{fig:postpub_governance_b}
  \end{subfigure}
  \caption{Post-publication learning: (a) Credit dynamics; (b) Long-term governance adaptation.}
  \label{fig:postpub_governance}
\end{figure*}

Figure~\ref{fig:postpub_governance} shows how delayed outcomes drive credit evolution and slower governance adaptation.
Author credit increases with sustained impact and decays with repeated low-impact outputs, while reviewer credit updates by calibration between ex ante assessments and ex post outcomes (Fig.~\ref{fig:postpub_governance}a).
Lagged feedback also induces gradual, bounded policy drift with inertia and saturation (Fig.~\ref{fig:postpub_governance}b), consistent with conservative long-horizon learning rather than reactive control.

% ------------------------------------------------------------
\subsection{Collusion Suppression and Mitigation Ablation}
\label{subsec:results_collusion}

\begin{figure*}[!htb]
  \centering
  \begin{subfigure}[t]{0.48\linewidth}
    \centering
    \includegraphics[width=\linewidth]{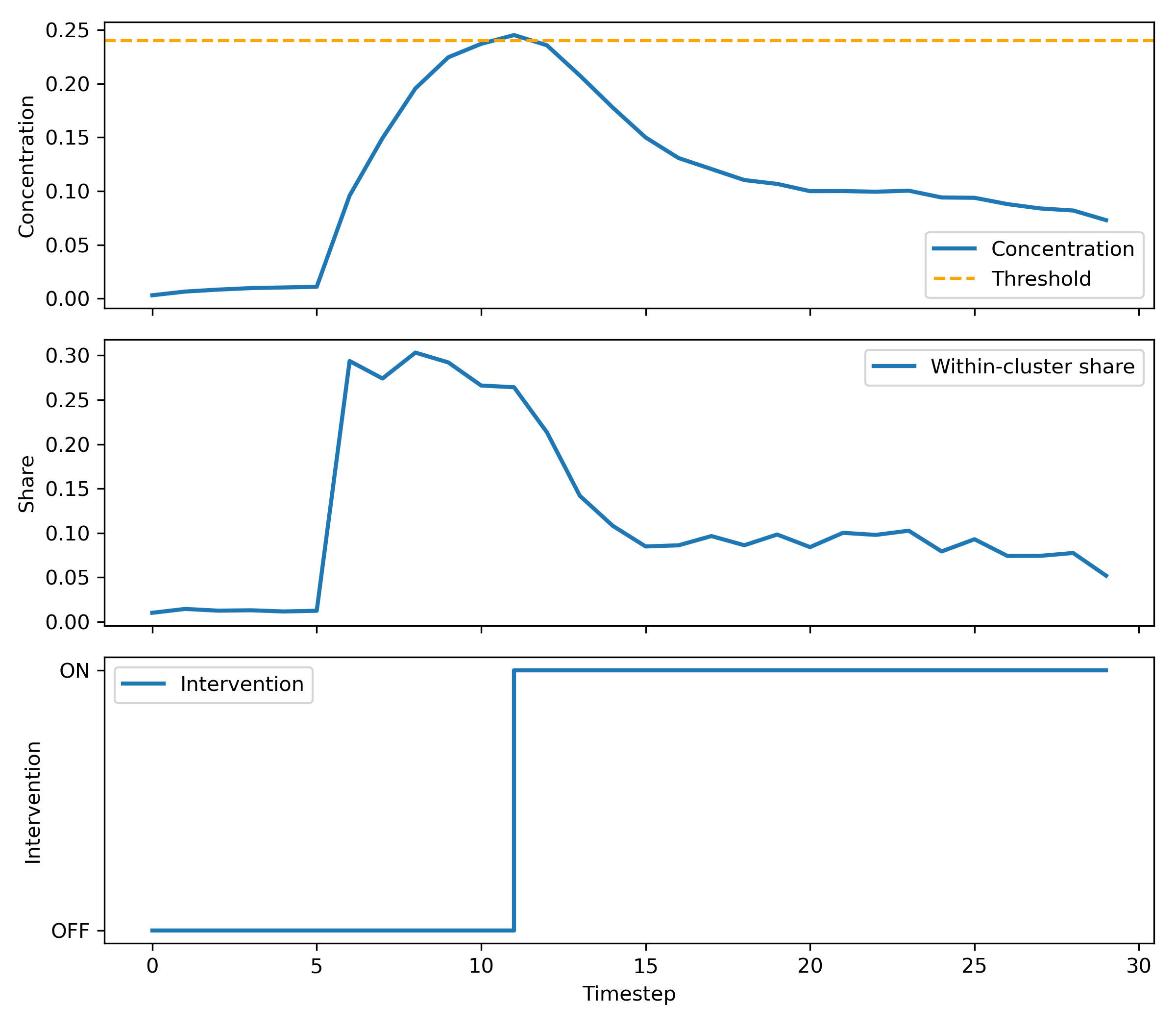}
    \caption{Mitigation enabled (representative run)}
  \end{subfigure}
  \hfill
  \begin{subfigure}[t]{0.48\linewidth}
    \centering
    \includegraphics[width=\linewidth]{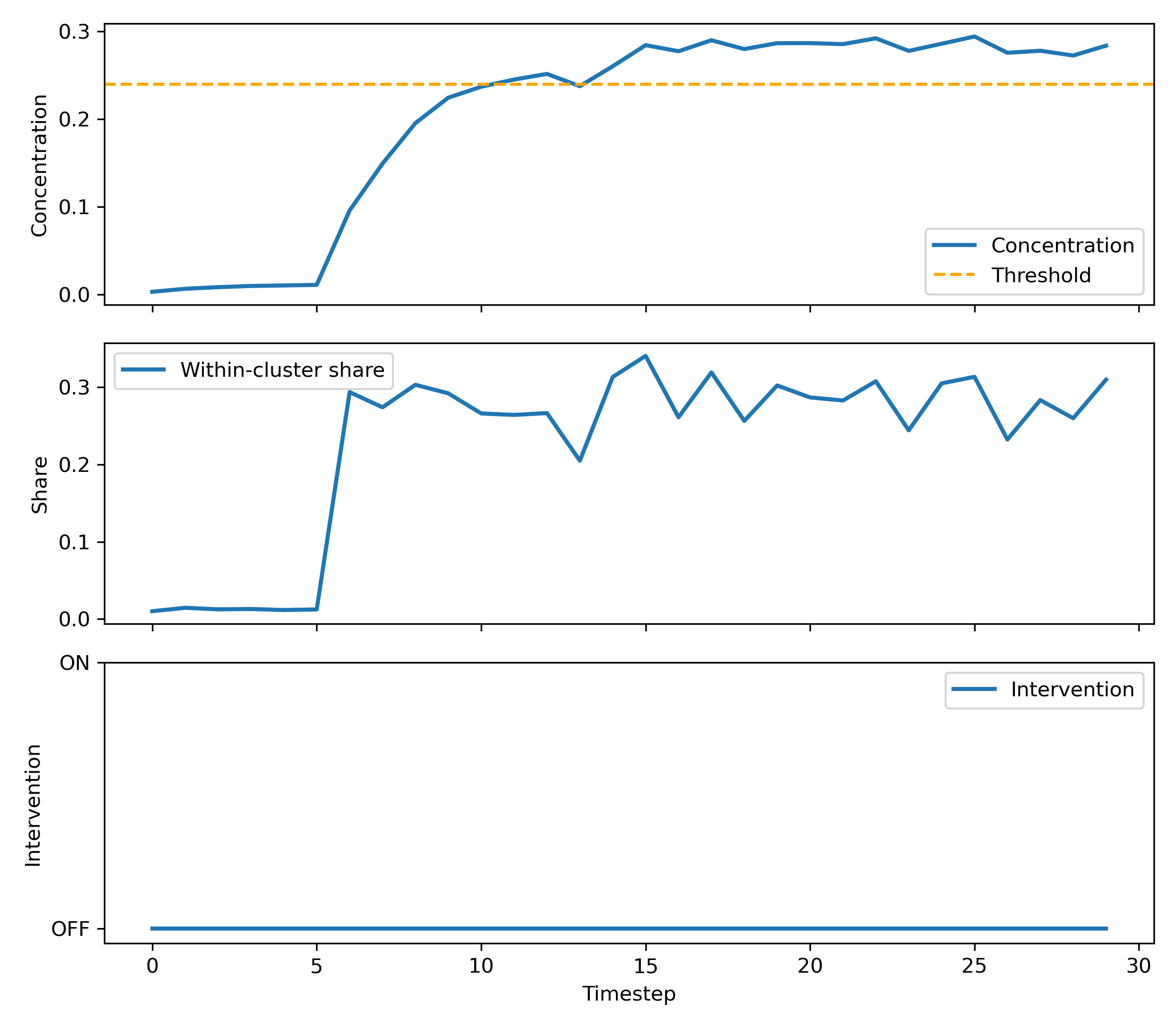}
    \caption{Mitigation disabled (representative run)}
  \end{subfigure}
  \caption{Collusion/cluster capture stress test: (a) Mitigation enabled; (b) Mitigation disabled.}
  \label{fig:collusion_capture}
\end{figure*}

Figure~\ref{fig:collusion_capture} evaluates a stylized collusion regime in which a coordinated reviewer subset increases within-cluster concentration.
We report multi-seed robustness over 10 random seeds.

\textbf{Mitigation enabled.}
Concentration reliably crosses the detection threshold and triggers a decentralization intervention, after which concentration decays.
Across seeds, the median maximum concentration is $0.244$ (range $0.240$--$0.265$) and the median final concentration is $0.073$ (range $0.061$--$0.094$).
Intervention activates at a median $t{=}11$ (range $10$--$13$).

\textbf{Mitigation disabled.}
In the no-mitigation ablation, intervention never activates and concentration remains elevated.
Across seeds, the median maximum concentration is $0.294$ (range $0.284$--$0.306$) and the median final concentration is $0.277$ (range $0.268$--$0.295$).
Peak within-cluster share is also higher without mitigation (median $0.327$, range $0.310$--$0.350$) than with mitigation (median $0.307$, range $0.296$--$0.338$), consistent with sustained cluster capture.

\section{Discussion}
\label{sec:discussion}

\subsection{ADAPT as a Governance Protocol Rather Than a Workflow Patch}
\label{subsec:disc_protocol}

ADAPT is best viewed as a governance-layer protocol that sits above individual editorial tasks.
In contrast to AI-assisted peer-review tools that optimize local workflow components (e.g., triage, reviewer recommendation, or critique drafting), ADAPT defines a closed-loop control mechanism that adapts \emph{policy parameters} in response to observable system signals.
This design targets system-level properties under stress, including bounded backlog, interpretable policy shifts, structured handling of uncertainty, and resilience to capture-like dynamics.
A key implication is that improvements at the task level do not automatically translate into stable system behavior when feedback is delayed and incentives are misaligned; policy-level control provides a principled mechanism to address these coupled effects over time.

\subsection{Human Authority and Accountable Control}
\label{subsec:disc_human_authority}

ADAPT separates \emph{decision authority} from \emph{signal processing and allocation}.
In a real deployment, accept/reject decisions and escalation judgments are human-ratified, while AI components are used to support scale (e.g., triage assistance, matching, and aggregation of review signals).
In our testbed, the decision step is implemented via a deterministic rule to enable controlled regime comparisons; ADAPT's contribution is evaluated through how policy parameters respond to stress and how those responses shape workload, disagreement, and concentration metrics.
Decentralization is operationalized through rotating roles and eligibility constraints informed by retrospective signals, which reduces long-run concentration of discretionary power and makes governance interventions attributable to logged signals.

\subsection{Implications for Scalable and Auditable Publishing}
\label{subsec:disc_implications}

ADAPT motivates a protocol-based notion of trust in scholarly communication.
Because governance actions are represented as explicit parameter updates and logged events, the system supports post hoc auditing of \emph{when} and \emph{why} interventions occurred, without exposing manuscript content, review text, or identities.
This property is compatible with community-run and open-access publishing models, where legitimacy depends on transparent rules and bounded discretion.
More broadly, ADAPT suggests that scalability and accountability can be improved simultaneously by constraining adaptation to interpretable policy knobs and by keeping governance actions externally verifiable.

\subsection{Limitations and Future Directions}
\label{subsec:disc_limitations}

\paragraph{Stylized simulation models.}
Our evaluation uses simplified stochastic models for arrivals, reviewer behavior, disagreement, and delayed impact.
These are intended for controlled stress testing, not field-specific realism.
A natural next step is calibration using conference or journal operational data, including discipline-specific arrival rates, turnaround distributions, and reviewer-load constraints.

\paragraph{Strategic and adaptive adversaries.}
We study a stylized collusion/cluster-capture regime and a mitigation ablation, but we do not model fully adaptive adversaries that respond strategically to governance policy.
Extending the simulator toward richer strategic behavior (e.g., adaptive collusion, reputation gaming, or manipulation of delayed impact signals) would provide stronger evidence for robustness.

\paragraph{Deployment and governance design choices.}
Practical deployment requires careful integration with submission platforms, privacy-preserving interfaces for escalation and audit, and explicit policies for conflict-of-interest handling.
Credit assignment and role eligibility must also be designed to reduce bias and limit metric gaming, particularly when delayed post-publication signals are used for learning.

\paragraph{Default instantiation versus journal-specific calibration.}
The present paper evaluates a default ADAPT simulator instantiation chosen for interpretability, controlled stress testing, and reproducible figure generation.
Accordingly, the current arrival model, review-noise model, time-cost model, collusion metric, and delayed impact proxy should be read as explicit default choices rather than claims of universal realism across journals.
A practical strength of this formulation is that these components can be replaced in a journal-specific adaptation while preserving the same governance interface, observable signals, and auditability structure.
This is also why we expose scenario configurations and a local reproducibility interface: a reader can perturb the default parameters, substitute alternative distributions or thresholds, and study how the same ADAPT control structure behaves under a different editorial setting.

\section{Conclusion}
\label{sec:conclusion}

We presented ADAPT, a decentralized and adaptive framework that reframes scholarly publishing as a closed-loop governance system.
ADAPT uses policy-level control driven by observable system signals, supports mixed human--AI participation without granting AI unilateral authority, incorporates delayed post-publication outcomes as retrospective learning signals, and provides auditable governance events under confidentiality constraints.
Across multiple stress regimes in a controlled simulation testbed, ADAPT exhibits bounded and interpretable responses, including backlog management under surges, structured handling of disagreement, and mitigation of a stylized cluster-capture scenario.
These results support the broader claim that scalable and trustworthy scholarly communication benefits from governance design that is adaptive, constrained, and auditable.

\section*{List of Abbreviations}
\begin{tabularx}{\linewidth}{@{}l >{\raggedright\arraybackslash}X@{}}
\textbf{Abbreviation} & \textbf{Full Term} \\
ADAPT & AI-Driven Decentralized Adaptive Publishing Testbed \\
AI & Artificial Intelligence \\
AP & Author pool (Authors $\mathcal{A}$) \\
DeSci & Decentralized Science \\
LLM & Large Language Model \\
\end{tabularx}

\section*{Declarations}
\addcontentsline{toc}{section}{Declarations}

\subsection*{Funding}
The authors received no external funding for this work.

\subsection*{Competing interests}
The authors declare that they have no competing interests.

\subsection*{Authors’ contributions}
Md Motaleb Hossen Manik led the development of ADAPT, implemented the simulation and experimental pipeline, and performed the evaluations.
Ge Wang supervised the project, guided the study design, and provided critical revisions to the manuscript.
All authors read and approved the final manuscript.

\subsection*{Acknowledgements}
Not applicable.

\subsection*{Availability of data and materials}
The code, configuration files, and data generated or analyzed during this study are available in the GitHub repository, \url{https://github.com/manikm-114/ADAPT_2}.

% \printbibliography

\end{document}